\documentclass[twocolumn,preprintnumbers,aps]{revtex4}
\usepackage{dcolumn}
\begin{document}
\title{Unconventional relaxation in antiferromagnetic CoRh$_2$O$_4$ nanoparticles}
\author{R.N. Bhowmik\footnote{Corresponding author(RNB):\\
e-mail: rabindranath.bhowmik@saha.ac.in\\
Present address: Department of Physics, Pondicherry University, R.V. Nagar, Kalapet, Pondicherry-605014} and R. Ranganathan\footnote{e-mail (RR): r.ranganathan@saha.ac.in}}
\affiliation{Experimental Condensed Matter Physics Division,\\
Saha Institute of Nuclear Physics, 1/AF, Bidhannagar, Calcutta 700064, India}

\begin{abstract}
Magnetic relaxation in antiferromagnetic CoRh$_2$O$_4$ nanoparticles is investigated at 2 K by cooling
the sample from a temperature (70 K) well above the antiferromagnetic ordering temperature at 27 K,
 following zero field cooled (ZFC) and field cooled (FC) process. In ZFC process, the sample at 2 K
is subsequently followed by magnetic field on and off sequences, whereas in FC process the cooling
field is made off during measurement of remanent magnetization as a function of time. The experiments suggest an unconventional relaxation behaviour in the system, as an effect of increasing surface exchange anisotropy with decreasing the size of antiferromagnetic nanoparticles.

\end{abstract}
\maketitle

Spinel oxides belong to a special class of materials with formula unit AB$_2$O$_4$ \cite{Buschow},
where cations are occupied in A (tetrahedral) and B (octahedral) sublattices, respectively.
Recently, many unusual and interesting phenomena like magnetic quantum tunneling and
superparamagnetism \cite{Tejada} have been
observed in spinel nanoparticles. These phenomena, owing to their tremendous applications in
nanoscience and technology, are drawing increasing attention to the research community.
The other notable phenomenon under current interest is the geometrical frustration effect.
The change of degeneracy and topology of the
antiferromagnetic ground state (Neel order) of a geometrically frustrated system
has shown various kinds of non-conventional magnetic
ordering \cite{Petro,Villain,Henley,Richert}. One such phenomenon is quantum spin fluctuation
effect, where the spins do not order and remain in a "collective paramagnetic state" down to
zero temperature. Some of the spinel oxides due to their typical lattice structure have
shown geometrical frustration effect \cite{gmf,Ueda}. Geometrical frustration effect has been
observed in those antiferromagnetic spinels where magnetic moments
are occupied either in B or in A sublattice. For example, ZnFe$_2$O$_4$ is
a spinel compound(structure: (Zn$^{2+}$)$_A$[Fe$^{3+}_{2}$]$_B$O$_4$) where
magnetic ions (Fe$^{3+}$) are confined in the B sublattice alone. On the other hand,
CoRh$_2$O$_4$ is another spinel (structure: (Co$^{2+}$)$_A$[Rh$^{3+}_{2}$]$_B$O$_4$) where
magnetic ions (Co$^{2+}$) are confined in the A sublattice alone.
Note that A-O-A (intra A sublattice) superexchange interactions are generally believed to be
smaller in comparison with B-O-B (intra B sublattice) superexchange
interactions. However, CoRh$_2$O$_4$ (A-O-A) and ZnFe$_2$O$_4$ (B-O-B) both are antiferromagnet, and
T$_N$ $\approx$ 27 K of CoRh$_2$O$_4$ is larger than T$_N$ $\approx$ 10 K of ZnFe$_2$O$_4$
\cite{Buschow}. In view of the earlier experimental results \cite{PRBCoRh2O4}, we believed that
CoRh$_2$O$_4$ can provide an alternative antiferromagnetic spinel for studying the geometrical
frustration effect.
We have seen that cubic spinel structure as well as
antiferromagnetic ordering temperature at 27 K of the bulk sample are also retained
in the CoRh$_2$O$_4$ nanoparticles. The unalternation of T$_N$ excludes the possibility of site exchange
between A sublattice (Co$^{2+}$) and B sublattice (Rh$^{3+}$) ions in CoRh$_2$O$_4$ nanoparticles.
The nanoparticles of CoRh$_2$O$_4$ (size down to 16 nm) were obtained by mechanical
milling of bulk sample. Mechanical milling is one
of the most convenient methods which can controll the surface spin ordering by contributing
some strain induced anisotropy \cite{RNBPRBstrain}.
The dc magnetization (M) as a function of time (t), temperature (T) and field (H)
were measured using SQUID (Quantum Design, USA) magnetometer. Before loading each sample in SQUID
magnetometer, zero field was checked by measuring the field dependence of magnetization of a paramagnetic
sample at 35 K.\\
We have analysed the zero field cooled magnetization [MZFC (T)] data above 50 K using Curie-Weiss law: $\chi$ = $\frac{C}{T-\theta_{A}}$, where C and $\theta_N$ have the usual meaning \cite{PRBCoRh2O4}. The analysis showed that effective paramagnetic moment ($\mu_{eff}$= (3kC/N)$^{1/2}$, where k is the Boltzman constant, N is the number of CoRh$_2$O$_4$ formula unit per gm of the sample) of the system increases with
the decrease of particle size. We also noted that $\mu_{eff}$ of the bulk sample is large
compared to $\mu_{eff}$ $\sim$ 4.60 $\mu_{B}$ for the Co$^{2+}$ moment alone.
Such an increase in Co-Rh
compound could be the effect of the spin-orbital coupling of 3d(Co)-4d(Rh) elements \cite{Zitum}.
It is found that the ratio of $\theta_A$/T$_N$ is always greater than 1 for nanoparticles. This is
a good indication of geometrical frustration effect in CoRh$_2$O$_4$ nanoparticles, as well as the increasing instability of antiferromagnetic ordering \cite{Palm} in surface spin configurations.
It is noted that the initial increase of $\theta_A$/T$_N$ ( -44.23 K/27 K for bulk, -42.80 K/27 K for 70 nm, -42.05 K/27 K for 50 nm, -41.84 K/27 K for 32 nm, -43.81 K/27 K for 19 nm and -51 K/27 K for 16 nm)
with decreasing the particle size down to 36 nm is followed by the decrease on further lowering the
particle size. This suggests that magnetic instability is not monotonous with decreasing
the particle size of the present antiferromagnetic spinel. A better magnetic ordering, which is different from the typical antiferromagnetic order,
can be observed for smaller particles.
The magnetic instability in the nanoparticle samples
is also reflected by the observation of magnetic irreversibility between the temperature dependence of
zero field cooled magnetization (MZFC) and field cooled magnetization (MFC) below T$_N$, whereas the increasing magnetic order with decreasing the
particle size is confirmed from the M(H) data. Such observation is also consistent with the
prediction of disorder induced magnetic order in geometric frustrated system \cite{Richert} or
induced ferromagnetism in antiferromagnetic nanoparticle \cite{Neel}.
The other notable feature consistent with disorder induced magnetic order is that magnetization
below T$_N$ had shown a systematic enhancement
with decreasing the particle size. A scaling analysis suggested
that such an increase of low temperature magnetization is a consequence of superparamagnetism
(collective paramagnetism), arising from the increasing number of frustrated surface (shell)
spins in nanoparticles.
The small irreversibility between FC and ZFC magnetization (MFC $>$ MZFC) below T$_N$ suggests
that a fraction of surface spins might be in blocking state and field cooling gives better
magnetic ordering of those blocked surface spins. However, we do not find any ferromagnetic hysteresis
loop in our nanoparticles, unlike other class of antiferromagnetic (AFMNP) nanoparticles
\cite{Makh,Barco,antidyn,Lierop} that showed exchange bias effect and the hysteresis loop like ferromagnetic materials. These unusual phenomena in AFMNP are the indication of strong inter-particle interactions, which has been reflected in
the blocking behaviour of nanoparticles below a characteristic temperature,
known as average superparamagnetic blocking temperature (T$_B$) of the nanoparticle sample.
In such cases, the superparamagnetc behaviour is characterized by the shift of T$_{B}$
with applied magnetic field in zero field cooled dc magnetization or with applied frequency in
ac susceptibility measurements. Since the blocking temperature, associated with a peak in
zero field cooled magnetization, is not observed in our AFMNP down to 2 K, an alternative approach is
essential for the understanding of surface superparamagnetism in antiferromagnetic nanoparticles,
like we are having.\\
The present work has probed the relaxation experiments to understand the
magnetic instability, essentially related to the surface spin dynamics, in
CoRh$_2$O$_4$ nanoparticles. The understanding could be relevant to
clarify certain issues like surface spin configuration, competitive effect between the
shell (surface) and core spins in antiferromagnetic nanoparticles.
In the relaxation experiments, samples are zero field cooled from 100 K
to the measurement temperature (T$_m$) at 2 K. After cooling, the samples are equilibrated
at 2 K in the absence of external magnetic field for the waiting time (t$_w$) $\sim$ 100 seconds.
Then external magnetic
field (H) = 100 Oe is applied, which is stabilized within 60 seconds.
The time dependence of the magnetization is recorded
for the next (t$_{on}$) 2700 seconds (definded as ON state). Then magnetic field is switched
off and the magnetic field is stabilised to zero value
within 60 seconds. However, the measurement of M(t) is continued for the next
(t$_{off}$) 2700 seconds in the absence of magnetic field (defined as OFF state).
We have shown the experimental data for two selected nanoparticle samples in Fig. 1.
The M(t) data during ON state
and OFF state are also separately shown for all the samples in Fig. 2 and Fig. 3a, respectively.
It is worthy to mention that relaxation behaviour is not observed for bulk sample. This is
consistent with the long range antiferromagnetic order of the sample. On the other hand, relaxation
behaviour is observed in all nanoparticle samples. We have noted a cross over in the magnetic relaxation behaviour depending on the size of particles. The experimental data show an unusual decrease of
magnetization with the increase of time, {\it i.e. negative magnetization growth with time}, during
the field ON state for samples with particle size down to $\sim$ 19 nm. Such magnetic relaxation behaviour
in the presence of magnetic field is unusual in view of the conventional domain growth, specially applicable for spin glass or superparamagnetic systems.
On further lowering the particle size, the system shows the as usual increase of magnetization with
time in the presence of magnetic field, {\it i.e. positive magnetization growth with time} (as seen for
sample with particle size $\sim$ 16 nm). Now, look at the M(t) data during field OFF state of
the measurement. Fig. 3a shows that the isothermal remanent magnetization M(t) during OFF state
decreases with the increase of time.
Interestingly, that decrease is becoming slower with time for the smaller particles. We have understood
such a slow decrease of M(t) during OFF state by comparing it with the convensional time decay of
field cooled remanent magnetization for 16 nm sample. The time dependence of remanent magnetization at
2 K is preceeded by 100 Oe field cooling from 70 K to 2 K (rate 5 K/minute) and subsequently, waited
there for 100 seconds before reducing the field to zero value and starts the recording. The remanent magnetization M(t) data is normalized by the first
point M(i) during OFF state, where i = 10 sec. and 2800 sec. for FC and ZFC process (during first 2700 sec. field is ON), respectively. The normalized
M(t)/M(i) data are shown in Fig. 3b.  We have seen that M(i) $\sim$
9.9$\times$ 10$^{-4}$ emu/g for FC-OFF is larger than M(i) $\sim$
2.0 $\times$ 10$^{-4}$ emu/g for ZFC-OFF process, and the magnitude of
M(t) is larger throughout the time (0-2800 seconds) after switing off the
cooling field in FC process in comparison with the OFF state (2800-5500 seconds) in ZFC process. Fig. 3b shows that the rate of decay of remanent magnetization
is more slower in FC process than that of the OFF state in ZFC process.
One would expect the establishment of a metastable magnetic order in the
surface spin configurations during the field cooling process from 70 K ($>$ T$_N$ $\sim$ 27 K) down to 2 K. The increase of interaction networks among the
surface spins can give rise a larger field cooled remanent magnetization, whereas the rate of decay of remanent magnetization is related to the strength of interactions among the spins, resulting in a slower decay for increasing interactions.

Most of the reports explained the
superparamagnetic behaviour in nanoparticle system by assuming the spin flips between spin up ($\theta_{0}$ = 0$^0$) and spin down ($\theta_{0}$ = 180$^0$) states, taking into account the anisotropy barrier E = - KVcos$^2$ $\theta_{0}$ \cite{Barco}.
This may not be a realistic model for nanoparticle system, also pointed out by other group \cite{spm}.
One would expect a distribution of canting angle ($\theta$) between surface spin configurations.
Consequently, we can expect a distribution of energy minima as a function of spin canting
angle (0$^0$ $<$ $\theta$ $<$ 180$^0$). The spin can relax in any one of the modified energy
minima, determined by local configurations.
The observed magnetic relaxation of CoRh$_2$O$_4$ nanoparticles can be explained by
the core-shell model \cite{PRBCoRh2O4}, which we proposed by taking into account the variation of
surface spin canting angle with particle size. We have considered that interaction between core spins are antiferromagnetic, whereas
interaction between core-shell spins are distorted antiferromagnetic.
On the otherhand, interaction between shell spins are effectively
superparamagnetic. When the particle size is reduced,
some of the A-O-A superexchange bonds become frustrated.
These frustrated bonds will create exchange anisotropy field at the
interfacial surface.
The decrease of magnetization with time in the presence of constant magnetic field suggests
a competition between AFM exchange interactions of core (bulk) and
surface anisotropy field (unidirectional) at the surface spins.
The competitive effect of the interactions is explained with the help of a schematic diagram (Fig. 4).
For spin vectors, the effective magnetic field (along + z axis) H$_{eff}$ =
H$_{ex}$ + H$_{san}$ + H$_{appl}$ + $\Delta$, where H$_{ex}$ is the Heisenberg
exchange interactions field (negative for antiferromagnetic ordering), H$_{san}$ is the surface
exchange anisotropy field, H$_{appl}$ is the applied external field, and $\Delta$ is small contribution from other sources.
For typical bulk AFM, H$_{san}$ = 0, H$_{ex}$
$>>$ H$_{appl}$ (100 Oe in the present experiment), and perfect AFM ordering compensates the magnetization vectors (Fig. 4a).
The distortion of AFM order in nanoparticle favours canting
between shell spins. The canting (0$^0$ $<$ $\theta$ $<$ 180$^0$) between two nearest neighbour
spins gives rise the resultant magnetization vector ($\mu$),
which makes angle $\phi$ (0 $<$ $\phi$ $<$ 90$^0$) with
the z direction (Fig. 4b). The magnetization (M) is the effective component of ($\mu$) along +z
direction, {\it i.e.} M = $\mu_{z}$. Remember that resultant surface exchange anisotropy field may
not be along the
+ z axis. We have assumed that H$_{san}$ is the effective component of surface exchange anisotropy field
along + z axis.
The effect of H$_{san}$ is to increase the surface magnetization
\cite{surfaceaniso,Takano} by decreasing
$\phi$, whereas H$_{ex}$ tries to compensate magnetization vectors ($\mu$ = 0). This establishes a
non-equilibrium local magnetic ordeing of the surface spins, before attaining an equilibrium
magnetic state M$_{eq}$ (H,T,t). The effective magnetization M(T) of (larger size) nanoparticle at
a constatnt temperature below T$_N$ may be larger, due to the contributions from H$_{san}$, than the corresponding bulk sample. At the same time, there is also
a finite probability for the shell spins of nanoparticle, under temperature and field equilibrium for a longer observation time, of restoring the bulk
M(T) state, if the surface spin configuration of nanoparticle sample does not drastically differ from
the bulk (180$^0$) configuration \cite{antidyn}. The time evolution of the local non-equilibrium
magnetization may not have any significant effect, because of the very
small magnitude, on the observed value of thermal activated magnetization at constant external magnetic field (100 Oe). However, such time evolution (relaxation dynamics of shell spin configurations) can effectively show the decrease of
magnetization even in the presence of external magnetic field, as long as H$_{ex}$ $>>$ H$_{san}$.
Now, consider the case where H$_{san}$ begins to dominate over H$_{ex}$ for antiferromagnetic
nanoparticles below a critical size (Fig. 4c). Here, antiferromagnetic field H$_{ex}$ is
effectively reduced in magnitude (schematically shown in Fig. 4 in Ref. \cite{PRBCoRh2O4} and the effect
is already found in many antiferromagnetic nanoparticles \cite{Makh,Barco,antidyn,Lierop}).
The effective decrease of antiferromagnetic exchange interactions results in the further decrease of
$\phi$ for smaller particles. Consequently, the probability of restoring the bulk M(T) state becomes
less. On the other hand, the local magnetic ordering of the shell spins in the presence
of external magnetic field favours the establishment of a new metastable state
where M(T) data increases with the increase of time.

In summary, Bulk (core) magnetization is a stable antiferromagnetic state, whereas
magnetization of nanoparticles can be considered as an distorted antiferromagnetic state.
At thermal equilibrium the core spins of nanoparticles will try to align the neighbouring shell
spins in antiferromagnetic manner, which results in a finite probability of decreasing the
surface magnetization with time. On the other hand, surface anisotropy field will try to
give rise a preferential induced magnetic ordering to the shell spins, which results in the
increase of surface magnetization with time. When the particle size is large
(i.e. surface anisotropy field is small), there is a greater probability for the
nanoparticle sample of restoring the equilibrium magnetization M$_{eq}$ of bulk sample at
that temperature. On decreasing the particle size, such probability of restoring the bulk
magnetic state becomes less, and the sample is locally ordered to a metastable state
where magnetization increases with time. This demonstrate the evidence of an
unconventional relaxation process in the system. The experimental results also provided an alternative
approach to identify qualitatively the increasing surface exchange anisotropy with decreasing the size of antiferromagnetic nanoparticle.

\end{document}